\documentclass{elsart}
\usepackage{amsmath}

\usepackage{graphicx}

\newcommand{\be}{\begin{eqnarray}}
\newcommand{\ee}{\end{eqnarray}}
\begin{document} 

\begin{frontmatter}

\title{Statistical production of antikaon nuclear bound states in heavy
ion collisions}

\author[gsi]{A. Andronic},
\author[gsi]{P. Braun-Munzinger},
\author[gsi,wro]{K. Redlich}

\address[gsi]{Gesellschaft f\"ur Schwerionenforschung, Darmstadt, Germany}
\address[wro]{Institute of Theoretical Physics University of Wroclaw,
PL--50204 Wroc\l aw, Poland }


\begin{abstract}
Recently it was conjectured  that the strongly attractive
antikaon--nucleon potential  can result in  the formation of
antikaon nuclear bound states. We  discuss the formation of such
states as possible residues in heavy ion collisions. In this
context, we calculate the excitation functions of single- and
double-$K^-$ clusters in terms of the statistical thermal model.
We show that, if such objects indeed exist, then, in heavy ion
collisions, the single-$K^-$ clusters are most abundantly produced
at present SIS energies, while the double-$K^-$ clusters show a
pronounced maximum yield in the energy domain of the future
accelerator at GSI. This is a direct consequence of: i) the
baryonic dominance in low energy heavy ion collisions and the
large baryonic content of the antikaonic bound states; ii) the
strong energy dependence of strangeness production at low
energies.
 The production yields of double-strange clusters is compared with
that of double strange baryons. It is shown that at SIS energy
there is a linear scaling relation  of the  $\Xi^-/K^+$ with
$K^+/p$ yields ratio.

\end{abstract}

\begin{keyword}
bound kaonic clusters
\sep statistical model
\PACS 25.75.-q, 25.75.Dw, 24.10.Pa
\end{keyword}

\end{frontmatter}


\section{Introduction}

Heavy ion collisions provide an experimental environment to study
the properties of nuclear matter under conditions of high energy
density well above its nuclear saturation. The observed particle
production yields and spectra were  shown to be sensitive to the
collective  and possible medium effects in a thermal system created
in heavy ion collisions \cite{r2,r1,r3,r4,r5,r6}.
 In low energy collisions at SIS accelerator a particular role
 has been attributed to strange mesons production.
 There are theoretical expectations that kaons and antikaons should
experience different interactions in high density nuclear matter
\cite{r4,new,kn,volker,s0,s1,s2}. 

Results obtained within a coupled-channel approach based on the chiral 
SU(3) effective Lagrangian show \cite{s1,s2} that $K^+$ undergoes repulsive
interaction whereas $K^-$ undergoes attractive interaction with
nucleons.
 However, it is not established yet  how strong is the
antikaon attraction with the nuclear matter. The earlier study
\cite{friedman1,friedman11} of kaonic atoms has indicated a very
strong attraction (150--200 MeV). More recent results
\cite{baca,friedman1s} suggest  relatively shallow (50--60 MeV)
potential. Also, self consistent calculations \cite{lutz} of
in--medium potential for $\bar K$ that include  its  in-medium
mass-shift find a shallow antikaon optical potential not exceeding
30 MeV. The effects of higher partial waves \cite{wa} and the pion
dressing \cite{pion} are also  modifying  the strength of
in-medium antikaon interactions. 
The self-consistent coupled channel calculations performed so far suggest 
that at normal nuclear matter density the in-medium  potential for antikaons
is at most 80 MeV. 

 The attractive in-medium optical potential for antikaons and
repulsive  interaction of kaons should, in general, result in  a distinct 
difference between $K^-$ and $K^+$ production cross section and flow pattern 
in heavy ion collisions.
 Indeed, the recent results of the KaoS \cite{kaos} and FOPI
\cite{fopi} collaborations indicate such a difference. Within the
context of dynamical transport \cite{tr1,tr2,tr3,tr4}
and statistical  model \cite{kmid0,kmid1} calculations
of strangeness production at SIS energies it was argued that the
observed properties of kaon and antikaon production can be to a
large extent understood if an in medium kaon--nucleon potential is
taken into account.

It was realized recently that, 
in spite of the strong absorptivity of the K$^-$-nucleus potential, 
sufficiently narrow deeply bound kaonic atom states could exist irrespective
of how attractive the K$^-$-nucleus potential is \cite{friedman}. 
In addition, the topic of strangeness at
low energies has gained an additional facet. Based on
phenomenologically constructed $K^-$-nucleon interactions, it was
argued that an additional consequence of a strongly attractive
$K^-$-nucleon potential could  be the formation of deeply bound
nuclear $K^-$ states \cite{kk1,kkd,kkh}. Such states should
exhibit central nuclear densities which exceed by 4--9 times the
normal nuclear density. They are also characterized by large
binding energies $(E_K\simeq 100$ MeV)\footnote{This value
is larger than the upper limit expected in self consistent
coupled channel calculations.} and  widths of 13--40 MeV
\cite{kk1,kkd,friedman2}. In Table~\ref{tab1} we summarize the
predicted $K^-$ clusters and their basic physical properties \cite{kkh}.

\begin{table}[htb]
\caption{The properties of the predicted $K^-$ clusters \cite{kkh}:
isospin $(I,I_z)$, spin-parity ($J^{\pi}$), mass ($Mc^2$), binding energy
($E_K$), width ($\Gamma_K$), nuclear density in the center of the system 
($\rho (0)$) and radius ($R_{\rm rms}$).}
\label{tab1}
\begin{tabular}{l|ccccccc}
$K^-$ -- cluster  & $(I,I_z)$ & $J^{\pi}$ & $Mc^2$ & $E_K$  &
$\Gamma_K$ & $\rho (0)$ & $R_{\rm rms}$  \\
                           &                    &                   &
[MeV]   & [MeV]    & [MeV]                  & [fm$^{-3}$] & [fm] \\
\hline $pK^-$ ($\Lambda(1405)$)  &$(0,0)$             & $(1/2)^-$  &1407
& 27     &
40     &  0.59     &   0.45      \\
$ppK^-$          & $(1/2,1/2)$ & $0^-$        & 2322  & 48     &
61
   & 0.52     &   0.99      \\
$pppK^-$       & $(1,1)$        & $(3/2)^+$  &3122  & 186   &   13
&  1.56     &    0.81     \\
$ppnK^-$       & $(0,0)$         & $(1/2)^-$  &3152  & 170   &
21
& 1.50   &   0.72       \\
$ppnK^-$       & $(1,0)$        & $(3/2)^+$  &3118  & 190   &   13
&  1.56     &    0.81     \\
$pnnK^-$       & $(1,-1)$        & $(3/2)^+$  &3117  & 191   &
13
&  1.56     &    0.81     \\
$ppppK^-$     & $(3/2,3/2)$  & $ 0^-$        & 4171- 90 & 75+90
&  20
  & 1.68        & 0.95  \\
$pppnK^-$     & $(3/2,1/2)$   & $0^-$       &4135-90   &113+90  &
20
  & 1.29 & 0.97\\
$ppnnK^-$     & $(3/2,-1/2)$ & $0^-$        &4135-90 & 114+90  &
20
&         & 1.12  \\
$ppK^-K^-$    & $(0,0)$         & $0^+$      & 2747      & 117
& 35 &  &  \\
$pppK^-K^-$ & $(1/2,1/2)$  & $(3/2)^+$ &3580-180        & 220+180
   &      &  &   \\
$ppnK^-K^-$ & $(1/2,-1/2)$  & $(3/2)^+$ &3582-180     & 221+180
& 37  & 2.97  & 0.69  \\
$pppnK^-K^-$ & $(1,0)$        & $0^+$       & 4511-180    &
230+180   &
61    & 2.33  & 0.73  \\
\end{tabular}
\end{table}

There are already some experimental indications \cite{kkd,kek1}
that $K^-$ nuclei could indeed be formed. An experiment carried
out at KEK has found \cite{kek1}  a peak in a neutron missing
momentum spectrum from the $^4He$(stopped $K^-, n$) reaction,
indicating a candidate for the predicted bound state of $ppnK^-$.
More recently a similar signal has been observed \cite{kek1} in a
proton missing spectrum indicating a narrow  $(I=1)$ state with  a
mass of 3117 MeV which could correspond to $pnnK^-$ cluster.

The $K^-$ nuclei, if they exist, could be not only populated by
direct reactions but could appear in any system  if the $K^-$ is
immersed in high  density nuclear environment. The dense medium
produced in heavy ion collisions could be a favorable environment
for the formation of such nuclei \cite{kkh}. Thus, the $K^-$
clusters may be possibly found as residues of relativistic heavy
ion collisions. This is particularly the case, as will be also
argued in this paper,  for the SIS energy range where antikaons
are produced in the high baryon density environment. Furthermore
the temperature reached at SIS energy is of the order or lower
than the  expected binding energy of $K^-$--nuclear  clusters.
Consequently, if such $K^-$ states are produced in heavy ion
collisions they cannot easily be destroyed by rescattering with
the surrounding thermal medium.

On the other hand it was argued \cite{wa,koch1} that, in
the heavy ion environment,  the structure of the
antikaonic potential could  be modified.  In heavy ion collisions
antikaons are produced with finite momentum, thus  formation of
$K^-$--nuclear  clusters is determined by the strength of the
potential at finite, instead of vanishing momenta. 
Theoretical studies \cite{wa,koch1} showed a non-trivial momentum
dependence of the potential which influences both its real and
imaginary part. There is in general a decrease of the strength of
the antikaon interaction in nuclear matter at finite momentum.

At high density \cite{mat} and/or temperature the spectral
function of antikaons is theoretically  expected to be so strongly
modified that the antikaon is not anymore a quasi-particle but rather 
a broad state with  temperature and/or
density-dependent width. In such a case it is rather difficult
to consider the binding of this broad state  with the surrounding
nucleons to form a bound nuclear cluster.  In ref. \cite{koch1} it was
even argued that the effect of scattering of antikaons with finite
momentum in a hot medium can wash out the attractive nature of the
potential. If this is  indeed the case, then the production of
$K^-$--nuclear clusters could be suppressed in heavy ion
collisions or their formation restricted  to the final stage,
close  to freezeout. We have to stress, however, that a detailed 
understanding of in-medium antikaon properties is at present 
not available.

In this paper we assume that kaonic  bound states can be  produced
in heavy ion collisions and discuss the production yields of
single- and double-$K^-$ clusters at chemical freezeout  in terms
of the statistical model. We compare the production probabilities
of such objects relative to those expected for $\Lambda$ or $\Xi^-$
strange particles in terms of the collision energy and system size
dependence.\footnote{The present calculations superseed the
preliminary ones shown in ref. \cite{aa-pbm}, for which an
incorrect assignment of the quantum numbers was done.}

\section{Outline of the statistical model}

The statistical model has been shown to be well suitable to describe
different particle and light nuclear cluster  yields
 \cite{r2,r10,r11,r12,r80}
obtained in heavy ion collisions in a broad energy range from SIS,
through, AGS, SPS and RHIC. To formulate the model we use the statistical
operator of hadron resonance gas HRG \cite{r2}. The resonance
contribution is of crucial importance to  describe effectively the
strong interactions of produced hadrons in the vicinity of
the chemical freezeout. On the other hand, resonances provide an
essential contribution to light particle yields. In the standard
formulation all resonances with well established decay properties
are included in the partition function \cite{r2,r10,r11}.

In nucleus--nucleus collisions particle production is constrained
by the conservation laws.  Thus, modelling the partition function
one needs to implement the conservation of the baryon number,
electric charge and strangeness. The first two conservation laws
are usually included on the grand canonical level and are
controlled in the statistical operator through the corresponding
chemical potentials. Strangeness conservation must be,  however,
introduced exactly within the canonical ensemble \cite{can1} 
where it is not anymore controlled by the strange chemical
potential or corresponding fugacity parameter. This is
particularly the case if one considers strangeness production in
heavy ion collisions at SIS energy \cite{can2}. There, strange
particles and antiparticles are very rarely produced and are
strongly correlated in order to preserve strangeness conservation.
Consequently, the thermal phase space available for strangeness
production is suppressed \cite{can22,can5}. This suppression  is
effectively described by the exact strangeness conservation
through canonical formulation of the partition function. A
detailed description of particle production in terms of the
statistical model can be found in Ref. \cite{r2,can3}. In the
following we summarize the basic results which are required to
quantify the yields of strange $K^-$ -- clusters and also other
strange particles  produced in heavy ion collisions. We emphasize
that no medium modifications of kaon properties are considered
within the present model.

The canonical partition function  of a hadron resonance gas in a
thermal system with total strangeness $S$ is obtained \cite{can3}
as:
 \be
Z^C_{S=0}=e^{S_0}
\sum_{n=-\infty}^{\infty}\sum_{p=-\infty}^{\infty} a_{3}^{p}
a_{2}^{n} a_{1}^{{-2n-3p}} I_n(x_2) I_p(x_3) I_{-2n-3p}(x_1)
 ,
\label{eq7a}
 \ee
where
%
 $a_i= \sqrt{{S_i}/{S_{-i}}}$,  $x_i = 2V \sqrt{S_iS_{-i}}$
%
and $S_i$ is the sum  of all $Z^1_k$  partition functions
\be Z_k^1= {{g_k}\over {2\pi^2}} \, m_k^2 \, T \, K_2(m_k/T)  \,
\exp (B_k\mu_B+Q_k\mu_Q) \label{eq4a}
 \ee
for particle species $k$ carrying strangeness $S_k=i$, the baryon
number $B_k$ and electric charge $Q_k$. The $I_n(x)$ in
(\ref{eq7a}) are the modified Bessel functions.

The density  $n_k^s$ of particle $k$ having the strangeness $s$ is
obtained from Eq.~(\ref{eq7a}) as:
 \be
n_{k}^s={{Z^1_{k}}\over {Z_{S=0}^C}}
\sum_{n=-\infty}^{\infty}\sum_{p=-\infty}^{\infty} a_{3}^{p}
a_{2}^{n}
 a_{1}^{{-2n-3p- s}} I_n(x_2) I_p(x_3) I_{-2n-3p- s}(x_1)
 .
\label{eq9a}
 \ee

In the canonical formulation of strangeness conservation the
density of strange particles is explicitly volume dependent
through the arguments $x_i$ of   the Bessel functions  in Eq.
(\ref{eq9a}). In the application of Eq. (\ref{eq9a}) to the
description of  particle production in heavy ion collisions this
volume parameter was interpreted as the strangeness correlation
volume which depends on  the number of participants \cite{can5}.
For large $V$ and for high enough temperature such that all
$x_i>>1$ the canonical result (\ref{eq9a}) is converging to its GC
value where strangeness conservation is controlled by the
corresponding chemical potential \cite{r2}. Obviously, in the GC
limit particle densities are not any more dependent on the volume
parameter. In heavy ion collisions the GC approximation was  found
to be adequate for energies beyond AGS \cite{r2}. For lower
collision energies, in particular for SIS, the suppression
due to canonical effects can even exceed an order of magnitude for
the yields of $S=\pm 1$ strange particles. The antikaonic
clusters are strangeness $S=-1$ or $S=-2$ objects, thus in the
context of a thermal model their production yields is to be
suppressed due to the exact strangeness conservation constraints.

Note that the above formulations do not explicitely include
finite widths. However, in the numerical realization of the model,
all widths of resonances and of the kaonic bound states are included
following the methods described in ref. \cite{r2}.

\section{Statistical production of $K^-$ nuclei in heavy ion
collisions}

In the thermal model the strange particle density (\ref{eq9a})
depends in general on four independent parameters: the
temperature, the charge and baryon chemical potential and the
correlation volume. Two of these parameters are fixed through the
initial conditions. In A--A collisions the isospin asymmetry in
the initial state fixes the charge chemical potential whereas the
correlation volume parameter $V$ is taken to scale with A \cite{can2,can5}
(see the discussion in Section \ref{k-sis}).
Thus, only the temperature $T$ and the baryon chemical
potential $\mu_B$ are left as independent parameters. In heavy ion
collision the temperature and baryon chemical potential  are the
parameters which characterize the properties of the collisions
fireball at chemical freezeout. These parameters are specific to
a given collision energy and can also vary with the number of
participating nucleons in the collision.

\subsection{Energy dependence}

 To quantify the production yields of $K^-$ clusters in
heavy ion collisions for different collision energies  we adopt
here the $\sqrt s_{NN}$ dependence of $T$ and $\mu_B$ following the
phenomenological chemical freezeout conditions of  fixed energy
per particle,  $<E>/<N>=1$ GeV \cite{1gev}. This condition was
shown to be  consistent with  chemical particles freezeout in
central Au+Au or Pb+Pb  collisions obtained in the  energy range
from SIS up to RHIC.
The thermal parameters extracted from the midrapidity data
and for  energies beyond the top AGS are also well consistent with
the fixed net baryon density as a chemical freezeout condition
\cite{r80}.

In nucleus--nucleus collisions  the  particle yield rather than
its density is an observable. In the thermal model the strange
particle yield is obtained from Eq. (\ref{eq9a}) when multiplying
the result by the fireball volume at the chemical freezeout.
As the fireball volume is not a directly measurable
quantity\footnote{The fireball volume can be infered from HBT
studies \cite{ceres}.},
to avoid an extra parameter we normalize the $K^-$ cluster yield
to the $\Lambda$ yield. In this case the ratio depends only on
the chemical freezeout values of $T$ and $\mu_B$.

\begin{figure}[htb]
\vspace{-1.cm}
\begin{tabular}{cc}
\begin{minipage}{.48\textwidth}
 \includegraphics[width=1.\textwidth]{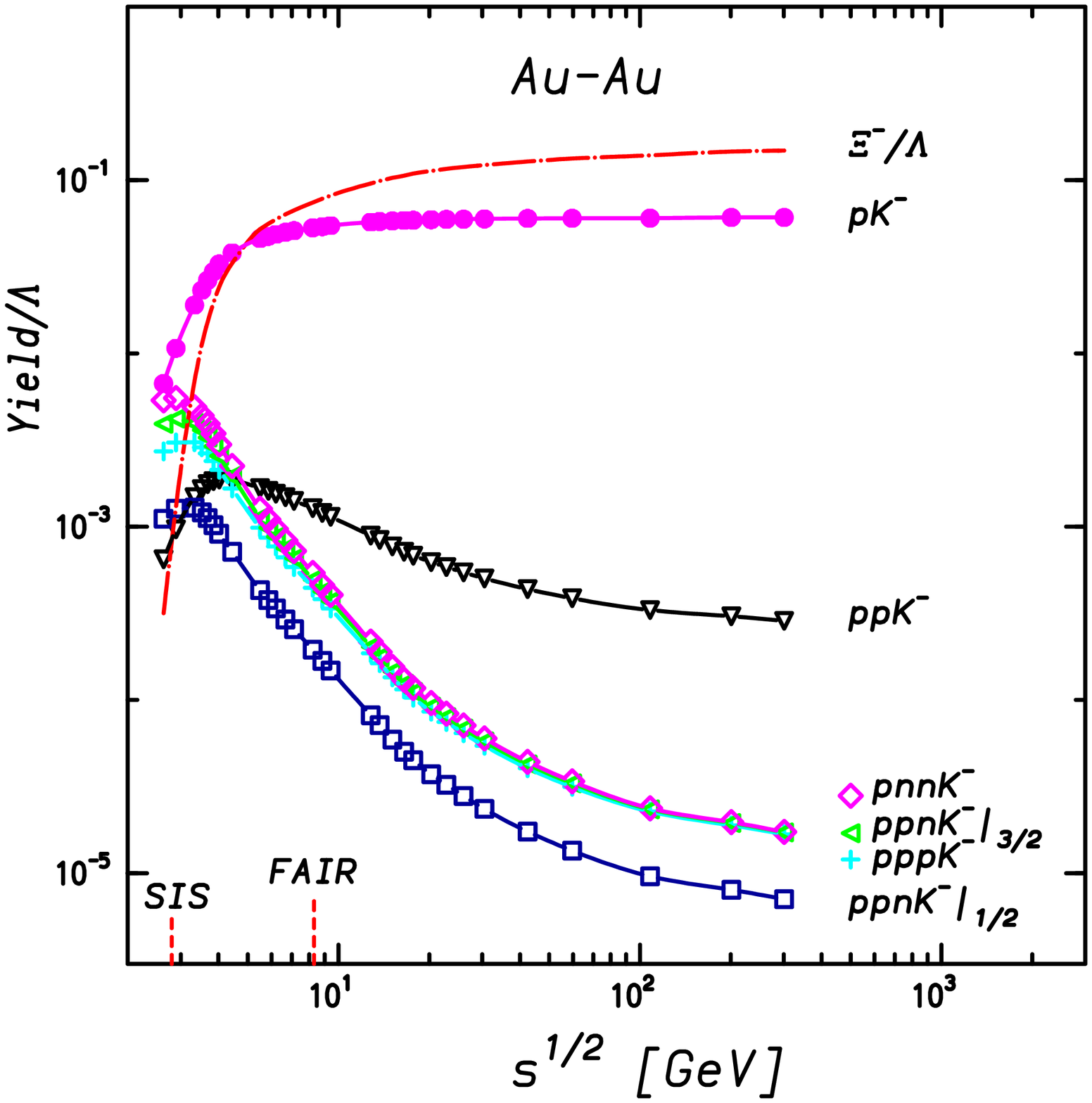}
\end{minipage}  & \begin{minipage} {.48\textwidth}
 \includegraphics[width=1.\textwidth]{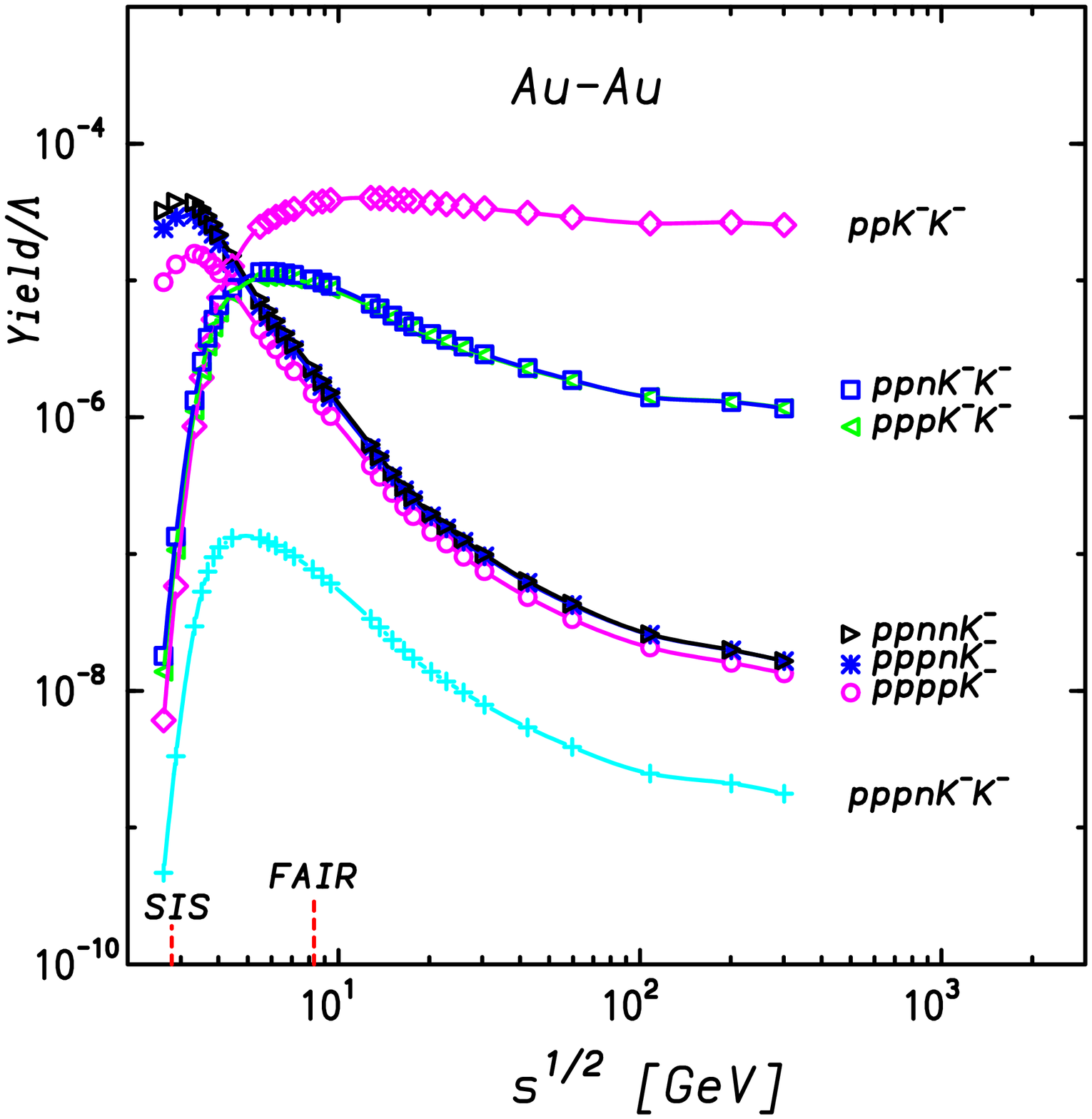}
\end{minipage}
\end{tabular}
\vspace{-1.cm} \caption{Excitation function of the yields of
antikaon bound states relative to $\Lambda$ yields. The
calculations are done along the freezeout curve of fixed $E/N=1$
GeV \cite{1gev}. For comparison, also show in the left panel is
the $\Xi^-/\Lambda$ yield ratio. The vertical lines indicate
the upper energies accessible in experiments at SIS and FAIR.}
\label{fig1}
\end{figure}


In Fig.~\ref{fig1} we show the resulting excitation
function for $S=-1$ and $S=-2$  antikaonic  clusters calculated in
the thermal model described above. The required parameters characterizing
the physical properties of these objects were taken from  Table 1. In
Fig.~\ref{fig1} the $K^-$--cluster/$\Lambda$   ratio is compared to 
the ratio $\Xi^-/\Lambda$. For $S=-1$ clusters the   yield/$\Lambda$ is 
seen in Fig.~\ref{fig1} to be a decreasing function of $\sqrt s$. This
is due to an interplay between $T$ and $\mu_B$ variation with the
collision energy. The increase of $T$ with  $\sqrt s$ should in
general enhance the yield/$\Lambda$ ratio. However, due to the
simultaneous drop in $\mu_B$ the yield/$\Lambda$ ratio decreases
with $\sqrt s$. The largest production rate of $S=-1$ and $B>1$
clusters is expected  at SIS energy  due to the large  $\mu_B\sim
0.7-0.82$ GeV reached at the  chemical freezeout. For $B=1$ state
a drop in temperature at SIS energy is not anymore compensated by
an increase in $\mu_B$, resulting  in a depletion of  the
yield/$\Lambda$ ratio. There is also a quite  steep decrease of the
$S=-2$ cluster yields towards SIS energy as seen in Fig.~\ref{fig1}.
This is due to: i) the large cluster mass, which through the Boltzmann
factor in Eq. (\ref{eq4a}) reduces the  thermal particle phase space;
ii) the strangeness suppression  effect. The strangeness suppression
factor is common for all particles carrying  the same strange quantum
numbers, thus it is cancelled out in the yield/$\Lambda$ ratio for
$S=-1$ clusters. For $S=-2$ states this is not any more the case as
the strangeness suppression is increasing with the strangeness
content of the particle \cite{can5}.

At SIS energy the yield of $S=-$1 low lying kaonic clusters is
larger than $\Xi^-$ yield. This is due to the  large baryon number
of these  objects which through the chemical potential increases
the thermal phase space of $K^-$ nuclei beyond that accessible for
$\Xi^-$. There is also an additional  suppression of $\Xi^-$ due
to the strangeness conservation as discussed above. For heavier
$S=-1$ clusters and for $S=-2$ states (see Fig.~\ref{fig1}) the production
probability is already an order of magnitude lower than for $\Xi^-$.

The excitation functions for antikaonic bound states shown in
Fig.~\ref{fig1} were obtained  for central Au+Au collisions.
Changing the colliding system is differently affecting the $S=-1$ and $S=-2$
yields. For $S=-1$ clusters the Yield/$\Lambda$ ratio is essentially
independent on the number of participants. This is due to the
cancellation of the correlation volume dependent factors in the
above  yields ratio. Some changes in $S=-1$  Yield/$\Lambda$ ratio
with A in A--A collisions are to be expected due to the different
isospin asymmetry in the initial state. The isospin effect appears
e.g. in Fig.~\ref{fig1} as the difference between $pppK^-$ and $pnnK^-$
yield at low $\sqrt s$.

The results of Fig.~\ref{fig1} indicate clearly that the SIS energy
range is the most appropriate to search for single-$K^-$ clusters
in heavy ion collisions.
Since such an experimental program is underway at the SIS accelerator
\cite{fopik}, we further focus on the model predictions in this 
energy range.
On the other hand, the yield of double-$K^-$ clusters is maximal
in the energy range of the future accelerator planned at GSI \cite{fair},
where it can be addressed within the CBM experiment \cite{cbm}.

\subsection{ Kaonic nuclear clusters at SIS energy} \label{k-sis}

The freezeout temperature in A--A collisions at SIS energies is so
low ($T\simeq 50-70$MeV) that all arguments $x_i$ in (\ref{eq9a})
are less than unity. In the limit of $x_i<<1$ it is sufficient to
take only the term with $n=p=0$ in Eq. (\ref{eq9a}). In addition,
expanding the Bessel functions $I_i(x)$ for $x_i\to 0$ it is also
sufficient  to  consider only the leading term. Within the above
approximations  the density $n_i^s$ of particle $i$ carrying
strangeness $s$ is obtained from Eq. (\ref{eq9a}) as

\be
n_i^s\simeq Z^1_i{{V^{| s|}}\over {{| s|}!}}
\times
 \left\{ \begin{array}{ll}
                          (S_{ -1})^{| s|}~, &  s>0 ~\\
                        (S_{+1})^{| s|}~, &  s < 0~
                    \end{array} \right. \label{eq10a} 
\ee
where \be  S_{-1}\simeq
Z^1_\Lambda+Z^1_{\Sigma^0}+Z^1_{\Sigma^+}+Z^1_{\Sigma^-} ~~~~{\rm
and }~~~~~ S_{+1}\simeq Z^1_{K^+}+Z^1_{K^0}  \label{eq11a} \ee
with $Z^1_i$ defined as in Eq. (\ref{eq4a}).

Considering the  structure of  Eqs. ({\ref{eq10a}) and
(\ref{eq11a}) it is clear that strange particles appear  in pairs
to guarantee  the   total strangeness to be  exactly zero. In A--A
collisions at SIS energy the correlation volume parameter $V$
is supposed to scale with $A$ as $V\simeq A\, V_0$ with $V_0=(4/3)\pi
r_0^3$ with $r_0\simeq 1.1$ fm. Such parametrization of the
correlation volume together with the canonical description of
strangeness production   is consistent with Au--Au and Ni--Ni data
on  $K^+$ and $K^-$ production at SIS energies \cite{can2,sis1}.

The $\bar K$ nuclear clusters,  if produced during heavy ion
collisions, should follow the thermal model systematics.  The
cluster yields $\langle N\rangle_{S=-1}^c $ and
 $\langle N\rangle_{S=-2}^c $ carrying strangeness
$S=-1$ and $S=-2$ respectively and  normalized to the number of
 $\Lambda$ are found  from Eq. (\ref{eq10a}) as

\be
{{\langle N\rangle_{S=-1}^c}\over {\langle \Lambda\rangle}}\simeq
{{Z^1_{S=-1}}\over {Z^1_\Lambda+Z^1_{\Sigma_0}}}~,~~~~~~~
{{\langle N\rangle_{S=-2}^c}\over {\langle \Lambda\rangle}}\simeq
{1\over 2}\,V\,{{Z^1_{S=-2}}\over {Z^1_\Lambda+Z^1_{\Sigma_0}}}(
{Z^1_{K^+}+Z^1_{K_0}}),\label{eq12a}
\ee
where the  one particle partition functions  $Z^1_{S=-1}$ and
$Z^1_{S=-2}$ are  calculated  from Eq. (\ref{eq4a}) with the
corresponding cluster parameters taken from Table~\ref{tab1}.

\begin{figure}[htb]
\vspace{-1.cm}
\centering\includegraphics[width=.54\textwidth]{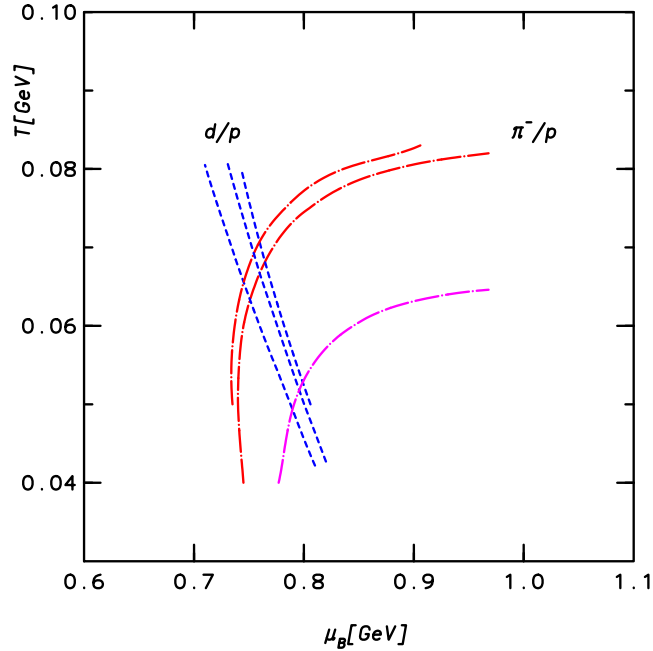}
\vspace{-1.cm}
\caption{The lines of constant $d/p=0.2, 0.26, 0.28$ (counted from left
to right) and $\pi^-/p=0.08, 0.17, 0.193$ (counted from bottom to top)
in the $(T,\mu_B)$ plane.}
\label{fig3}
\end{figure}

To quantify the  model predictions for the cluster yields in A--A
collisions at SIS energy one needs to specify the values of
thermal parameters at chemical freezeout. For a given colliding
system the relative production rate (\ref{eq12a}) depends  only on
$T$ and $\mu_B$. A very transparent method to pin down  these
parameters is illustrated in Fig.~\ref{fig3}. Considering the
experimental data on $d/p$ and $\pi/p$ ratio in the $T$--$\mu_B$
plane the freezeout point is determined as the crossing of
($d/p$=const.) and ($\pi/p$=const.) lines.\footnote{At SIS
energy the $d/p$ yields ratio was found \cite{r2,can2} to be
consistent with thermal model predictions when being
calculated  with the same parameters as all other ratios. 
This is why the $d/p$ ratio can be used to determine chemical
freezeout parameters in heavy ion collisions at SIS energies.}
For Au+Au collisions at $E_{lab}=1$ AGeV and in Ni+Ni collisions
at $E_{lab}=1.8$ AGeV the freezeout parameters are ($T\simeq
52$MeV, $\mu_B\simeq 822$ MeV) and ($T\simeq 70$MeV, $\mu_B\simeq
750$ MeV), respectively \cite{sis1}.
\begin{table}[htb]
\caption{Calculated yields for single-$K^-$ clusters for Au+Au
at 1 $A$GeVand Ni+Ni collisions at 1.8 $A$GeV.}
\label{tab2}
\begin{tabular}{l|l|c}
Yield/$\langle \Lambda\rangle$ &Au+Au 1 AGeV & Ni+Ni 1.8 AGeV   \\
\hline
$pK^-$            &0.399E-02              &   0.142E-01     \\
$ppK^-$          & 0.476E-03           &   0.129E-02       \\
$pppK^-$       &  0.309E-02                &  0.362E-02    \\
$ppnK^-|_{J^\pi=(1/2)^-}$       &  0.128E-02              &  0.128E-02  \\
$ppnK^-|_{J^\pi=(3/2)^+}$       &   0.483E-02                &  0.409E-02   \\
$pnnK^-$       &  0.713E-02               & 0.443E-02  \\
\end{tabular}
\end{table}

 Table~\ref{tab2}  summarizes the model results for the lower lying
$S=-1$ cluster yields in Au+Au and Ni+Ni collisions at  $E_{lab}=1$ AGeV
and $E_{lab}=1.8$ AGeV, respectively.
Comparing these results, it is clear that only
$pK^-$ and $ppK^-$  yields  are  sensitive to the collision
energy. The heavier states are only weakly affected by an increase
in $E_{lab}$  from 1 to 1.8 AGeV. This is rather a surprising
result as there is an essential change in the freezeout parameters
at these two colliding energies. In addition, due to the different
isospin asymmetry  in  Au and Ni nucleus there is also  a shift in
the charge chemical potential from $\mu_Q\simeq -19$ MeV  in
Au+Au to $\mu_Q\simeq 0.0$ MeV in Ni+Ni collisions.

The possible presence of kaonic nuclear clusters in heavy ion
collisions can in general be verified at the SIS accelerator within
the FOPI experimental program  dedicated  to study strangeness
production in Al+Al reactions at 2 AGeV. One way to extend the
predictions of the thermal model  to higher energy and
different colliding systems would be to extrapolate  the actual
values of the thermal parameters based on the previous systematics.
The dependence of freezeout parameters on the  system size at SIS
energy is, however, at present  still not well established. On the
other hand, $T$ and $\mu_B$ are not  direct observables in heavy ion
collisions. These thermal parameters can be directly
related with observables as shown in Fig.~\ref{fig3}. We express
the thermal parameters $(T, \mu_B)$ through the $d/p$ and $\pi/p$ ratio
and study the dependence of the relative production probability of
$\bar K$ clusters on the values of these ratios. We consider the
change of the cluster production rate with the $\pi/p$ ratio
calculated along a line of constant $d/p$ ratio as shown in
Fig.~\ref{fig3}.


\begin{figure}[htb]
\vspace{-1.cm}
\centering\includegraphics[width=.54\textwidth]{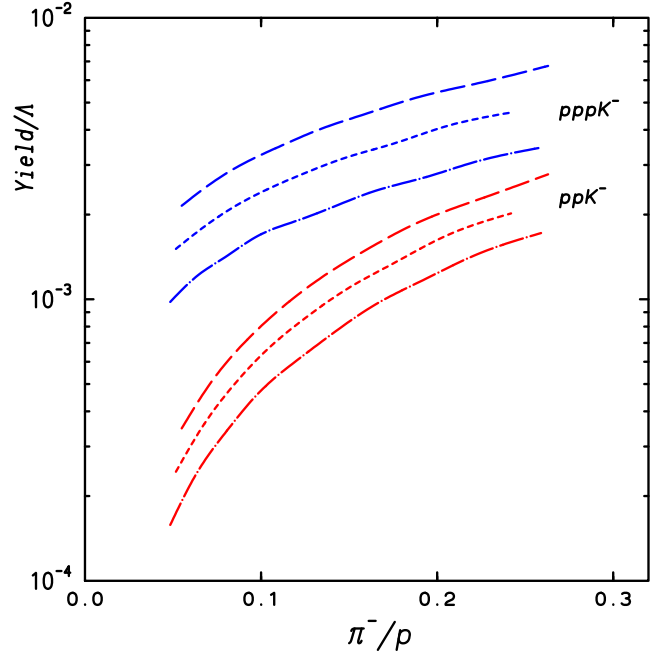}
\vspace{-1.cm}
\caption{The yield of $ppK^-$ and $pppK^-$ clusters  per $\Lambda$
yield as a function of the $\pi^-/p$ ratio. The calculations are done
for the  fixed $d/p$ ratio. For each cluster the short-dashed line
corresponds to $d/p=0.23$, the lower line to $d/p=0.20$ and the upper
line to $d/p=0.26$ respectively.}\label{fig4}
\end{figure}

In Fig.~\ref{fig4} the cluster yields are expressed as  a function of
the $\pi/p$ ratio for different values of  $d/p$. The $d/p$ yield
ratio is experimentally known to be a decreasing function of the
collision energy. In Ni+Ni collisions at $E_{lab}=1.8 $ AGeV the
$d/p\simeq 0.28$.  In  Fig.~\ref{fig4} we illustrate the $ppK^-$ and
$pppK^-$ relative yields  for three different values of
$d/p<0.28$. Having established the experimental value of $d/p$ and
the  corresponding $\pi/p$ ratio in Al+Al collisions at
$E_{lab}=2.0$ AGeV , the expected thermal model results for
cluster production probabilities can be obtained from Fig. 4.

\begin{figure}[htb]
\vspace{-1.cm}
\begin{tabular}{cc}
\begin{minipage}{.48\textwidth}
 \includegraphics[width=1.\textwidth]{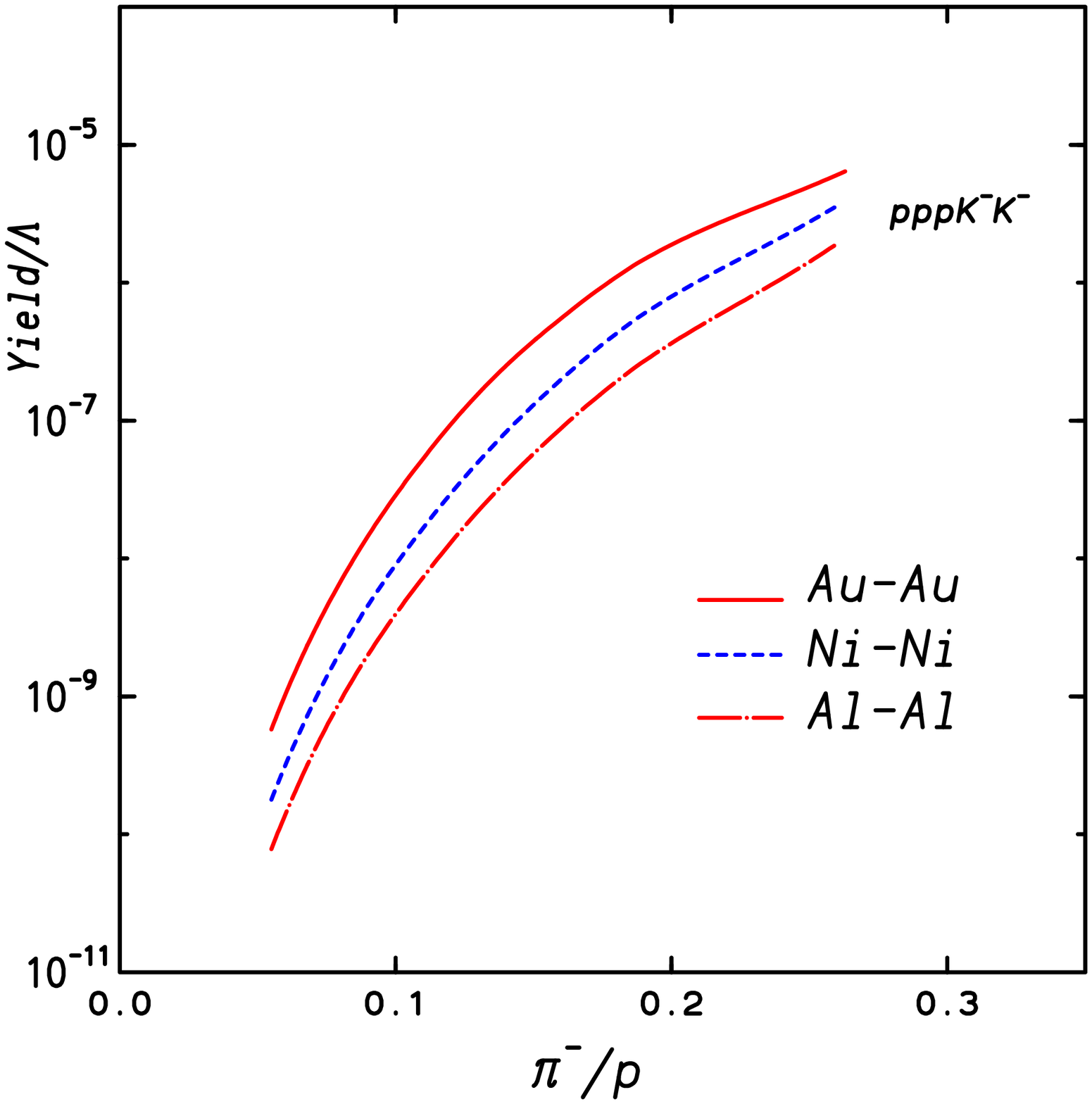}
\vspace{-1.5cm}
\caption{The yield of $pppK^-K^-$ per $\Lambda$ yield as a function of
$\pi^-/p$ ratio for Au+Au, Ni+Ni and Al+Al collisions. The calculation are
done for fixed $d/p=0.26$ yield ratio.}
\label{fig5}

\end{minipage}  & \begin{minipage} {.48\textwidth}
 \includegraphics[width=1.\textwidth]{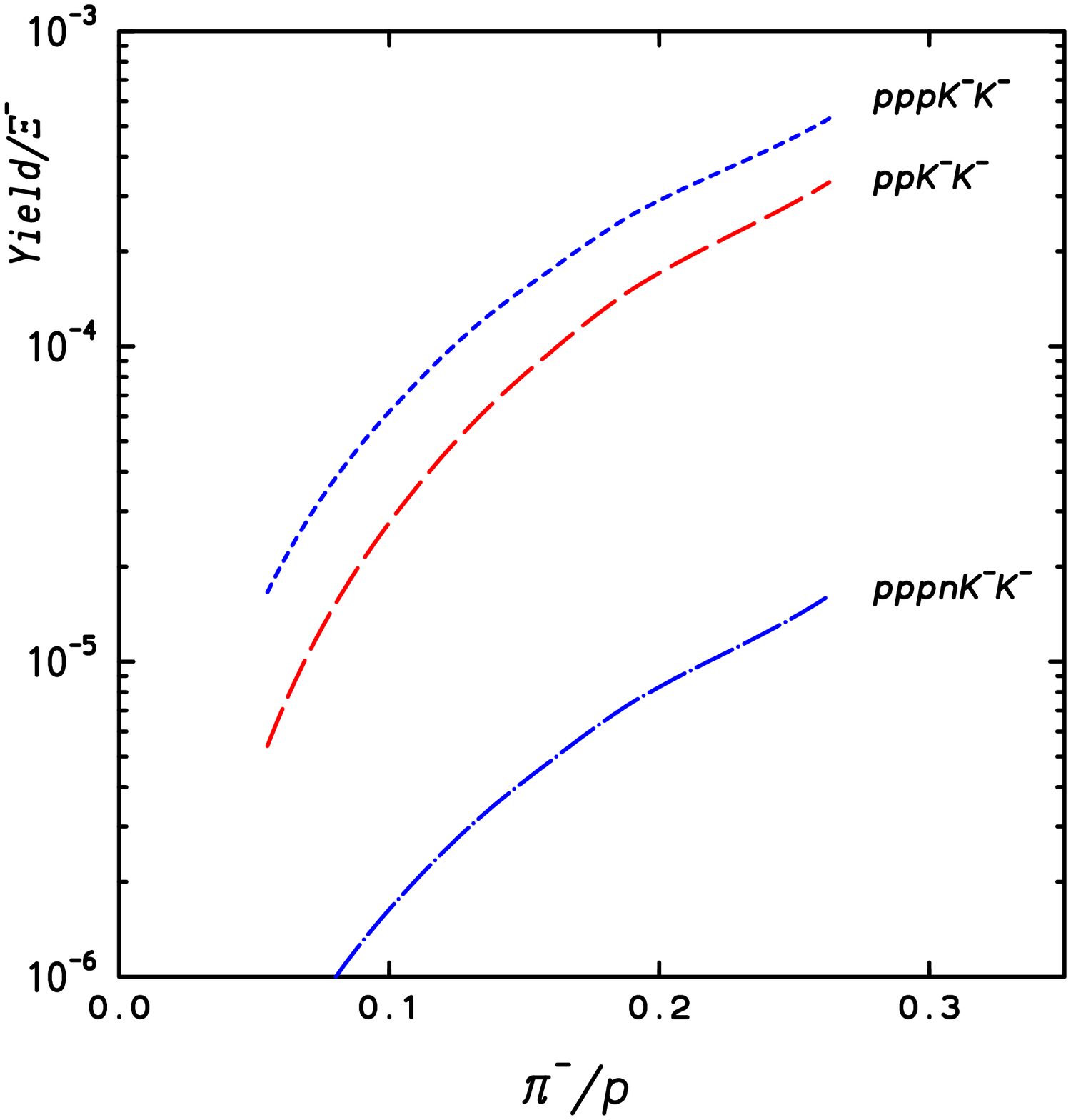}
\vspace{-1.5cm}
\caption{The yield of $ppK^-K^-$, $pppK^-K^-$ and $pppnK^-K^-$ clusters
per $\Xi^-$ yield as a function of  $\pi^-/p$ ratio. The calculations are
done for the fixed $d/p=0.26$ yields ratio.}
\label{fig6}
\end{minipage}
\end{tabular}
\end{figure}

The relative yields of double strange clusters are seen in Eq.
(\ref{eq12a}) to be  explicitly  dependent on the system size in
the initial state. In addition,  the system size dependence enters
through different isospin asymmetry and possible modification of
the $(T,\mu_B)$ freezeout parameters. In Fig.~\ref{fig5} we illustrate
the $A$--dependence of relative $pppK^-K^-$ cluster yield as a
function of $\pi/p$ ratio along the line of constant $d/p=0.26$.
For fixed $\pi/p$ ratio, the yields are dropping with decreasing
system size. For a sufficiently large $\pi/p$ value, that is at large
$\sqrt s$, the canonical suppression is less important and the
A--dependence of the production rate is negligible.


The explicit volume dependence seen in the
ratio of S=-2 cluster yields per $\Lambda$ yields arises simply
because these particles have  different strange quantum numbers.
It is clear that normalizing the S=-2 cluster yield to 
$\Xi^-$ yeild the V dependence is cancelled out. Indeed, in a
thermal model the $\langle \Xi^-\rangle/\langle \Lambda\rangle$
ratio is represented by the same expression as
$\langle N\rangle^c_{S=-2}/\langle \Lambda\rangle$ in Eq. (\ref{eq12a}),
but with the replacement of ${Z^1_{S=-2}}$ by $Z^1_{\Xi^-}$.
 Thus, the ratio
\be
{{\langle N\rangle^c_{S=-2}}\over {\langle \Xi^-\rangle}}\simeq
{{Z^1_{S=-2}}\over{Z^1_{\Xi-}}}
\ee
is determined by the corresponding ratio of the available
thermal particle phase space expressed through the single particle
partition functions.

In Fig.~\ref{fig6} the yield of lower lying S=-2 clusters, normalized to
the $\Xi^-$ yield,  is shown as a function of the $\pi^-/p$ ratio,
for fixed $d/p=0.26$. These relative yields are not anymore explicitely
volume dependent, thus should be the same for all colliding systems.
Some small system size dependence could appear, as already indicated,
due to a possible modification of chemical freezeout parameters and
an isospin asymmetry. However, these effects cannot  change the general
conclusion that the yield of the double-$K^-$ nuclear bound clusters
is by more than three orders of magnitude less abundant  than the yield 
of $\Xi^-$. Thus, such  states are rather  unlikely to be identified
in heavy ion collisions at the SIS energy \cite{fopik}.
As we have shown in Fig.~\ref{fig1}, the yield of double-$K^-$ clusters
steeply increases (by about three orders of magnitude) up to
$\sqrt{s}\simeq$5, where it becomes comparable to the single-$K^-$ clusters
at SIS energies. This makes their detection feasible at FAIR \cite{fair},
despite the larger expected combinatorial background.
The CBM experiment at FAIR is designed to study rare probes, 
including multistrange baryons \cite{cbm}. By employing a silicon detector 
to measure displaced vertices and due to the good particle identification
and momentum resolution, CBM should be able to detect a variety of kaonic 
clusters if they are produced at chemical freezeout. The unknown decay 
channels and branching ratios represent, however, further difficulties
for quantitatively assessing the experimental detection possibilities.

\section{The scaling relation of cascade production at SIS energy }

In the previous section we have emphasized the practical aspect to
look experimentally for the yield of the double-$K^-$ relative to $\Xi^-$.
However, the yield of $\Xi^-$ is presently not known in heavy ion
collisions in the SIS energy range.
To facilitate optimizing future measurements in this respect, we like
to emphasize interesting scaling properties of the relative production
probabilities  of $\Xi^-$ yield  normalized to $K^+$ or $\Lambda$ yield,
predicted within the statistical model.
This scaling is illustrated in Fig.~\ref{fig7} where the yield ratio
$\Xi^-/K^+$ is plotted as a function of the $K^+/p$ ratio.
The observed linear scaling of $\Xi^-/K^+$ ratio with $K^+/p$ is
independent of the system size.
In Fig.~\ref{fig7} we indicate the thermal model predictions of the
relative  $\Xi^-/K^+$ yield for different $K^+/p$ ratios
obtained at the  SIS energy.

\begin{figure}[htb]
\vspace{-1.3cm}
\centering\includegraphics[width=.56\textwidth]{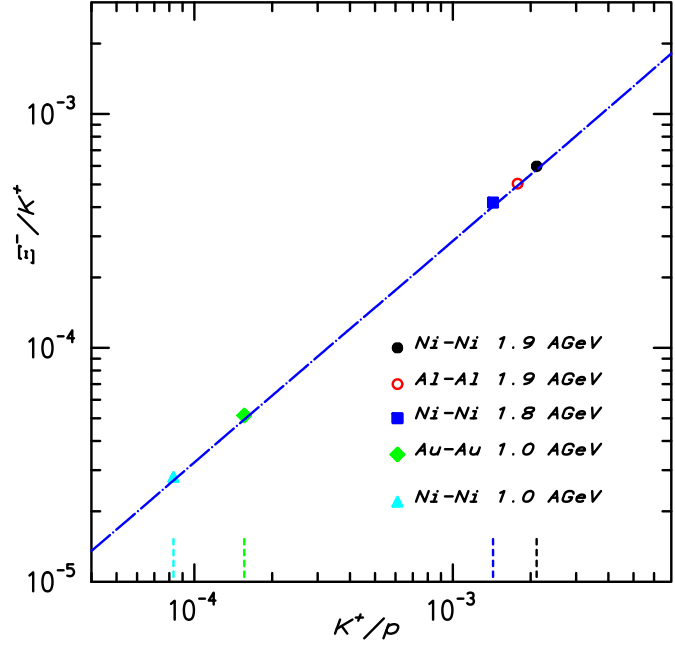}

\vspace{-1.5cm}
\caption{Scaling properties of $\Xi^-/K^+$ versus $K^+/p$ ratios.
The vertical dashed lines indicate the experimental values of $K^+/p$
ratio from KaoS and FOPI Collaborations \cite{kaos,fopi}
obtained in Au+Au and Ni+Ni collisions at two different incident
energies . The points on the scaling line are the expected results
of the $\Xi^-/K^+$ ratio for the corresponding  $K^+/p$ experimental
values. The Al+Al point is the prediction of the model.}
\label{fig7}
\end{figure}

The statistical model predicts that such a scaling should be
observed if the thermal conditions are such that the canonical
suppression effects are dominating the strangeness production
rates. The scaling has a more general, dynamical, origin and is a
consequence of the production mechanism of kaons, hyperons and
cascades at SIS energies.

The $\Lambda$ and $\Sigma$ hyperons are produced together with
kaons since this is energetically the most favorable way to
produce strange hadrons. On the other hand, antikaons are produced
below their threshold in binary N--N collisions, thus can be
produced only through  the strangeness-exchange reaction
\begin{equation}
\pi + Y \, \rightleftharpoons  \,  K^- + N. \label{eq1}
\end{equation}

If the rates for $K^-$ production are equal to those for   $K^-$
absorption, the  reaction  (\ref{eq1}) is in a chemical
equilibrium. In this case   the law-of-mass action is applicable
and leads to the following  relation between particle yields
\cite{sis1}
\begin{equation}
\frac{[\pi] \cdot [Y ]}{[K^-] \cdot [N]} \, = \, \kappa_1
,\label{eq2}
\end{equation}
with $Y=(\Lambda, \Sigma )$, $[x]$ being the multiplicity of
particle $x$ and with $\kappa$ being the chemical equilibration
constant~\cite{sis1}. On the other hand, the strangeness exchange
process

\begin{equation}
K^- + Y \, \rightleftharpoons  \,  \Xi^- + \pi \label{eq3}
\end{equation}
dominates the production of $\Xi^-$ in heavy  ion collisions at
SIS energies. Thus, here a detailed balance relation implies that

\begin{equation}
\frac{[K^-] \cdot [Y ]}{[\Xi^-] \cdot [\pi ]} \, = \, \kappa_2
\label{eq4}
\end{equation}
From  (\ref{eq2}) and (\ref{eq4}) one concludes that

\begin{equation}
\frac{[\Xi^-]}{[Y ]} \, = \, {1\over {\kappa_1\kappa_2}}
{{{[Y]}\over { [N]}}}\label{eq5}.
\end{equation}

This relation also holds for the reaction involving direct
absorption of $\Xi^-$ on proton, $\Xi^-+p \, \rightleftharpoons \,
\Lambda + \Lambda $: only if  such a process is  satisfying a
detailed balance relation.

To preserve strangeness conservation   the hyperons  are produced
together with $K^+$ and $K^0$ in equal rates,
$[Y]=[K^+]+[K^0]\simeq 2[K^+]$, thus from (\ref{eq5}) one can also
write
\begin{equation}
\frac{[\Xi^-]}{[K^+]} \, = \, \kappa\cdot {{{[K^+]}\over {
[N]}}}\label{eq6}.
\end{equation}

The above equation explains the linear scaling relation which
was obtained within the canonical thermal model, shown in Fig.~\ref{fig7}.
 The relation (\ref{eq5}) is  valid in the whole SIS
energy range for $0.8<E_{lab}<2$ A$\cdot$ GeV and is independent
of the atomic number of colliding nucleus  or centrality in A--A
collisions. Thus,  Eq. (\ref{eq5}) provides a very transparent
method to infer the yield of $\Xi^-$ from measurements of
the kaon and proton production rates at SIS.
 For higher collision energies the above scaling is expected to
be violated since the processes (\ref{eq1}) and (\ref{eq3}) are
not anymore dominating the production of $K^-$ and $\Xi^-$
hadrons.

Once the proportionality constant $\kappa$ in Eq. (\ref{eq5}) is
fixed, the scaling relation (\ref{eq5}) provides definite predictions
of the $\Xi^-$ production yield for different colliding systems.
 In general, $\kappa$ could be calculated within different models,
e.g. in dynamical transport models \cite{ko}, provided that such
models are preserving detailed balance relations for the
strangeness production. In the statistical model the scaling
(\ref{eq5}) appears naturally as a consequence of the canonical
strangeness conservation and chemical equilibration.

\section{Summary and conclusions}

The conjecture of the possible existence of deeply bound $K^-$
states has been studied for heavy ion collisions in the context
of the statistical thermal model.
Based on the analysis of the excitation function of the production
yield for various $K^-$ cluster species, we have concluded that
the maximum production probabilities of such objects appear at
present SIS energies for single-$K^-$ clusters and at the future
GSI accelerator in case of double-$K^-$ clusters.
This is a direct consequence of a large
baryonic density reached in A--A collisions at SIS
and of the strong energy dependence of strangeness production at
low energies.
We have discussed the production yields of single- and double-strange
clusters as a function of the system size and of the thermal composition
of the collision fireball created at SIS energies.
The model predictions on the production probabilities of antikaonic
nuclear states relative to the number of hyperons and cascades were
quantified.

We have argued that  the relative production yield of $\Xi^-/K^+$
in heavy ion collisions at SIS beam energies  should exhibit a
linear  scaling with the relative yield of $K^+/p$. Such a scaling
appears as a consequence of the  strangeness production mechanism
at subthreshold energies and the requirement of detailed balance
relations for all production reactions. In the statistical thermal
model the above scaling appears due to the  exact strangeness
conservation constraints.

\section*{ Acknowledgments}

We acknowledge stimulating   discussions with N. Herrmann, 
H. Oeschler, J. Wambach and T. Yamazaki.
K.R. acknowledges the support of the KBN under the grant 2P03 (06925).


\begin{thebibliography}{99}

\bibitem{r2} For  a recent review see e.g.: P. Braun-Munzinger, K. Redlich
and  J. Stachel, in {\it Quark Gluon Plasma 3  (Edts. R. Hwa and
X.-N. Wang)}; nucl-th/0304013.

\bibitem{r1}
E.L. Bratkovskaya, M. Bleicher, M. Reiter, S. Soff, H. St\"ocker,
M. van Leeuwen, S.A. Bass, and  W. Cassing, Phys. Rev. C69 (2004)
054907.

\bibitem{r3} U. Heinz, hep-ph/0407360.

\bibitem{r4} G.E. Brown and M. Rho, Phys. Rep. 363 (2002) 85.

\bibitem{r5} W. Cassing and H. Bratkovskaya, Phys. Rep. 308 (1999) 65.

\bibitem{r6} N. Xu, Prog. Part. Nucl. Phys. 53 (2004) 165.

\bibitem{new} C.B. Dover, and G.E. Walker, Phys. Rep. 89  (1982) 1.

\bibitem{kn} D.B. Kaplan, A.E. Nelson, Phys. Lett. B175  (1986) 57; 
A.E. Nelson, D.B. Kaplan, Phys. Lett. B192  (1987) 193.

\bibitem{volker} V. Koch, Phys. Lett. B337 (1994) 7.

\bibitem{s0} J. Schaffner-Bielich, J. Bondorf, and I. N. Mishustin,
Nucl. Phys. A625 (1997) 325.

\bibitem{s1} G.E. Brown, C.H. Lee, M. Rho, and V. Thorsson, Nucl.
Phys. A567 (1994) 937.

\bibitem{s2} T. Waas, N Kaiser, and W. Weise, Phys. Lett. B365 (1996)
12 and B379 (1996) 34; W. Weise, Nucl. Phys. A610 (1996) 35c.


\bibitem{friedman1}
E. Friedman, A. Gal, and C.J. Batty,  Nucl. Phys. A579  (1994) 518
and Phys. Lett. B308 (1993) 6.

\bibitem{friedman11} E. Friedman, A. Gal, J. Mares, and A. Cieply,
Phys. Rev. C60 (1999) 024314.

\bibitem{baca} A. Baca, C. Garcia--Recio, and J. Nieves,
Nucl. Phys. A673  (2000) 335.

\bibitem{friedman1s} A. Cieply, E. Friedman, A. Gal and J. Mares,
Nucl. Phys. A696 (2001) 173 and
Acta Phys. Polon. B35 (2004) 1011.

\bibitem{lutz} M.F.M. Lutz, Phys. Lett. B426  (1998) 12.

\bibitem{wa} L. Tolos, A. Ramos,  A. Polls, and T.T.S. Kuo,
Nucl. Phys. A690  (2001) 547.

\bibitem{pion}  L. Tolos, A. Ramos, and  A. Polls,
 Phys. Rev. C65  (2002) 054907; A. Ramos, and E. Oset,
Nucl. Phys. A671 (2000) 481.

\bibitem{kaos} A F\"orster, et al. (KaoS), Phys. Rev. Lett. 91 (2003)
152301. C. Sturm, et al. (KaoS), Phys. Rev. Lett. 86 (2001) 39. F.
Uhlig. et al. (KaoS), nucl-ex/0411021.

\bibitem{fopi} K. Wisniewski et al. (FOPI), Eur. Phys. J. A9 (2000)
515; P. Crochet et al. (FOPI), Phys. Lett. B486 (2000) 6; 
N. Herrmann et al. (FOPI), Acta Phys. Polon. B35 (2004) 1091.

\bibitem{tr1} J. Aichelin and C.M. Ko, Phys. Rev. Lett. 55 (1985) 2661; 
G.Q. Li and C.M. Ko, Phys. Rev. C54 (1996) R2159; 
J. Aichelin, Phys. Rep. 202 (1991) 233; 
C. Fuchs et al., Phys. Rev. Lett. 86 (2001) 1974; 
C. Hartnack and J. Aichelin, J. Phys. G28 (2002) 1649.

\bibitem{tr2} C. Hartnack, H. Oeschler and  J. Aichelin,
Phys. Rev. Lett. 90 (2003) 102302,  Phys. Rev. Lett. 94 (2004) 149903.

\bibitem{tr3} H. Oeschler, J. Phys. G27 (2001) 257.

\bibitem{tr4} A. Mishra, E. Bratkovskaya, J. Schaffner-Bielich, S.
Schramm, and H. St\"ocker, Phys. Rev. C70 (2004) 044904.

\bibitem{kmid0}
G.E. Brown, M. Rho, and C. Song, Nucl. Phys. A690  (2001) 184c and
Nucl. Phys. A698 (2002) 483c.

\bibitem{kmid1}
L. Tolos, A. Polls, A. Ramos, and  J. Schaffner-Bielich,
Nucl. Phys. A754 (2005) 356 and  Phys. Rev. C68  (2003) 024903.

\bibitem{friedman}
E. Friedman and A. Gal, Phys. Lett. B459  (1999) 43 and 
Nucl. Phys. A658 (1999) 345.

\bibitem{kk1} Y. Akaishi and T. Yamazaki, Phys. Rev. C65 (2002)
044005.
T. Yamazaki and Y. Akaishi, Phys. Lett. B535 (2002) 70.

\bibitem{kkd} T. Yamazaki, A. Dote, and Y. Akaishi, Proceedings of
the KIAS-APCTP International Symposium on Astro--Hadron Physics,
World Scientific (2004) 362.

\bibitem{kkh} T. Yamazaki, A. Dote, and Y. Akaishi, Phys. Lett. B587
(2004) 167. A. Dote, H. Horiuchi,  Y. Akaishi, and T. Yamazaki,
Phys. Lett. B590 (2004) 51.

\bibitem{friedman2}
J. Mares, E. Friedman, and A. Gal, Phys. Lett. B606 (2005) 295.

\bibitem{kek1} T. Suzuki et al., Nucl. Phys. A754  (2005) 375c.
T. Suzuki et al., Phys. Lett. B597 (2004) 263.

\bibitem{koch1} J. Schaffner-Bielich and V. Koch, Nucl. Phys.
A669  (2000) 153.

\bibitem{mat}
M.F.M. Lutz, E.E. Kolomeitsev, and C.L. Korpa,  J. Phys. G28
(2002) 1729 and references therein.

\bibitem{r10} P. Braun-Munzinger, D. Magestro, K. Redlich, and
J. Stachel, Phys. Lett. B518 (2001) 41 {and references therein};
F. Becattini, J. Cleymans, A. Keranen, E. Suhonen and K. Redlich,
Phys. Rev. C64 (2001) 024901;
J. Rafelski and J. Letessier, Acta Phys. Polon. B27 (1996) 1037;
F. Becattini, M. Gazdzicki, A. Keranen, J. Manninen, and R. Stock,
Phys. Rev. C69 (2004) 024905. 

\bibitem{r11}  P. Braun-Munzinger, J. Stachel, J.P. Wessels, and N. Xu,
Phys. Lett. B344 (1995)  43, {\it ibid.} B365 (1996) 1;
P. Braun-Munzinger and J. Stachel, Nucl. Phys. A606 (1996)  320;
P. Braun-Munzinger, I. Heppe, and J. Stachel, Phys. Lett. B465 (1999)
15.

\bibitem{r12} P. Braun-Munzinger and J. Stachel, J. Phys. G21 (1995) L17.

\bibitem{r80} P. Braun-Munzinger and J. Stachel, J. Phys. G28 (2002) 1971.

\bibitem{ceres} D. Adamova et al., Phys. Rev. Lett. 90 (2003) 022301.

\bibitem{can1}
R. Hagedorn, Thermodynamics of strong interactions, CERN Report
71-12 (1971). E.V. Shuryak, Phys. Lett. B42 (1972) 357; 
J. Rafelski and M. Danos, Phys. Lett. B97 (1980) 279; 
K. Redlich and L. Turko, Z. Phys. C5 (1980) 201;
R. Hagedorn and K. Redlich, Z. Phys. C27 (1985) 541.

\bibitem{can2}
J. Cleymans, H. Oeschler, and K. Redlich, Phys. Rev. C59 (1999)
1663 and Phys. Lett. B485 (2001) 27.

\bibitem{can22}  C.M. Ko, V. Koch, Z. Lin, K. Redlich, M. Stephanov, 
and X.N. Wang, Phys. Rev. Lett. 86 (2001) 5438.

\bibitem{can5}  A. Tounsi and K. Redlich,  Eur. Phys. J. C24 (2002)
529, J. Phys. G28 (2002) 2095; 
J. S. Hamieh, K. Redlich, and A. Tounsi, Phys. Lett. B486 (2000) 61,
J. Phys.  G27 (2001) 413.

\bibitem{aa-pbm} A. Andronic and P. Braun-Munzinger, hep-ph/0402291.

\bibitem{can3}
 P. Braun-Munzinger, J. Cleymans, H. Oeschler, and K. Redlich,
Nucl. Phys. A697 (2002) 902.

\bibitem{can4} K. Redlich and L. Turko, Z. Phys. B97 (1980) 279;
L. Turko, Phys. Lett. B104 (1981) 153.

\bibitem{1gev} J. Cleymans and K. Redlich, Phys. Rev. Lett. 81 (1998) 5284
and Phys. Rev. C60 (1999) 054908.

\bibitem{fopik} N. Herrmann et al. (FOPI), GSI Sc. Rep. 2004, p. 127;
L. Fabbietti et al. (FOPI), GSI Sc. Rep. 2004, p. 128. \\
{\tt http://www.gsi.de/informationen/wti/library/scientificreport2004/indexSR.html}.

\bibitem{fair} An International Accelerator Facility for Research with
Ions and Antiprotons (FAIR), {\tt
http://www.gsi.de/fair/index\_e.html}

\bibitem{cbm} The Compressed Baryonic Matter Experiment (CBM),\\
{\tt http://www.gsi.de/fair/experiments/CBM/index\_e.html}

\bibitem{sis1}
 J. Cleymans, A. F\"orster, H. Oeschler, K. Redlich, and  F.
 Uhlig, Phys. Lett. B603 (2004) 146.

\bibitem{sis2}
 J. Cleymans, H. Oeschler, and K. Redlich, Phys. Lett. B485 (2000) 27,
 J. Phys. G25 (1999) 281, and Pramana 60 (2002) 1039.

\bibitem{ko} Li-Wen Chen, C.M. Ko, and Y. Tzang, Phys. Lett. B584 (2004) 269; 
S. Pal, C.M. Ko, J.M. Alexander, P. Chung, and R.A. Lacey, 
Phys. Let. B595 (2004) 158. 


\end{thebibliography}
\end{document}